# Broad-Range Tuning of Ferroelectric Switching of La$_x$Bi$_{1-x}$FeO$_3$ Epitaxial Films via Digital Doping using Off-Axis Co-Sputtering

Katelyn Lazareno[1], Christopher Chae[2], Becky Haight[3], Shams Jabin[4], Rachel Steinhardt[5], John J. Plombon[5], Siddharth Rajan[4], Patrick M. Woodward[3], Jinwoo Hwang[2], and Fengyuan Yang[2]

[1]Department of Physics, The Ohio State University, Columbus, OH 43210, USA

[2]Department of Materials Science and Engineering, The Ohio State University, Columbus, OH 43210, USA

[3]Department of Chemistry and Biochemistry, The Ohio State University, Columbus, OH 43210, USA

[4]Department of Electrical and Computer Engineering, The Ohio State University, Columbus, OH 43210, USA

[5]Intel Foundry Technology Research, 2501 NE Century Blvd. Hillsboro, OR 97124, USA

## Abstract

To investigate the scope of ferroelectric behavior in La-substituted BiFeO$_3$ films, La$_x$Bi$_{1-x}$FeO$_3$ epitaxial films were synthesized using off-axis co-sputtering on SrTiO$_3$(001) and DyScO$_3$(110) substrates with a SrRuO$_3$ bottom electrode layer. A digital-doping deposition method was used to enable precise control and continuous tuning of La concentration in high-quality La$_x$Bi$_{1-x}$FeO$_3$ films across a wide range of $x$ = 0.05-0.60, which was systematically investigated using piezoresponse force microscopy. Robust and reversible out-of-plane ferroelectric switching has been observed up to $x$ = 0.35, while films with $x \geq 0.37$ exhibit no measurable ferroelectric behavior, indicating a sharp ferroelectric-to-paraelectric phase transition between $x$ = 0.35 and 0.37. This represents the highest reported La concentration in La$_x$Bi$_{1-x}$FeO$_3$ films that retains ferroelectric ordering, highlighting opportunities to engineer ferroelectric and multiferroic properties in complex oxide heterostructures.

BiFeO$_3$ (BFO) is a well-studied multiferroic material[1-3] that simultaneously exhibits strong ferroelectric (FE) and antiferromagnetic ordering with a small net canting moment in epitaxial films with ordering temperatures well above room temperature, making it a promising candidate for novel applications.[4-7] Despite these potential advantages, BiFeO$_3$ films present a few practical problems such as high leakage currents and large ferroelectric coercive fields, hindering its applicability and performance in devices.[8-10] One possible solution to overcome these issues is partial substitution of Bi with rare earth elements such as La at the A-site of the perovskite lattice, La$_x$Bi$_{1-x}$FeO$_3$, which has been shown to be effective in controlling the volatility of bismuth atoms and, thereby, suppressing leakage current densities while still maintaining its ferroelectric properties.[11-21] There, however, exists an upper limit to the doping concentration before the ferroelectric ordering vanishes. Since the ferroelectric behavior of the La$_x$Bi$_{1-x}$FeO$_3$ epitaxial films depends sensitively on the deposition techniques and growth parameters, the reported upper limit of La substitution in BiFeO$_3$ films that retain ferroelectric ordering varies considerably.[19, 22, 23] It is an interesting and important question for the research field to explore the highest La level in La$_x$Bi$_{1-x}$FeO$_3$ epitaxial films before room temperature ferroelectric polarization disappears.

In this work, we show the growth of a series of epitaxial La$_x$Bi$_{1-x}$FeO$_3$ films using "digital doping" by off-axis co-sputtering of BiFeO$_3$ and LaFeO$_3$ targets. We explore a wide range of La substitution regime in La$_x$Bi$_{1-x}$FeO$_3$ films with the La concentration $x$ from 0.05 up to 0.6, which is not typically investigated, to investigate the maximum La concentrations that can be introduced without loss of ferroelectric polarization. X-ray diffraction (XRD) characterization reveals high crystallinity of the La$_x$Bi$_{1-x}$FeO$_3$ films for the whole range of $x$ = 0.05 to 0.6. Ferroelectric properties of those films have been investigated using piezoresponse force microscopy (PFM), which demonstrates distinct polarization switching behavior up to 0.35 La substitution.

Stoichiometric, phase-pure BiFeO$_3$ and LaFeO$_3$ (LFO) powders were synthesized by solid-state reactions, which were then pressed into 2" diameter sputtering targets using our previously reported procedure for off-axis sputtering.[24] The BiFeO$_3$ and LaFeO$_3$ sputter sources are mounted horizontally with a 90° angle between their orientations. A single-crystal substrate is mounted at the center of a 2" substrate heater, which has a horizontal distance of 5.7 cm from and a vertical distance of 7.5 cm above the center of each sputtering target, resulting in an off-axis angle of 53°.

La$_x$Bi$_{1-x}$FeO$_3$ epitaxial films with $x$ = 0.05, 0.1, 0.15, 0.25, 0.3, 0.35, 0.37, and 0.6 were grown on a 10-nm SrRuO$_3$ (SRO) epitaxial layer on (001)-oriented SrTiO$_3$ (STO) and (110)-oriented DyScO$_3$ (DSO) substrates. SrTiO$_3$ (001) substrates (MTI) were treated by buffered hydrofluoric acid and annealed in air at 1050 °C for 2 hours while DyScO$_3$ (110) substrates were annealed in air at 1000 °C for 1 hour to form surface atomic terraces. A SrRuO$_3$ (10 nm) layer was first deposited on each substrate to serve as the bottom electrode by off-axis sputtering at 12 mTorr of Ar + 2% O$_2$ mixture using a DC sputtering current of 40 mA with a deposition rate of 5.7 Å/min at 450 °C and 500 °C on SrTiO$_3$ and DyScO$_3$, respectively.

La$_x$Bi$_{1-x}$FeO$_3$ epitaxial films were deposited at a substrate temperature of 650 °C using a mixture of Ar + 5% O$_2$ at a total gas pressure of 12 mTorr. These conditions are similar to what we used previously for growth of LaFeO$_3$ epitaxial thin films with excellent crystalline quality.[25-27] Here, the calibrated deposition rates for BiFeO$_3$ and LaFeO$_3$ are $r_{BFO}$ = 4.5 Å/min and $r_{LFO}$ = 3.2 Å/min using a radio-frequency (RF) sputtering power of 50 W and 45 W, respectively. In order to cover a broad range of La concentration from 0.05 to 0.6, if using conventional co-sputtering with both targets on for the whole growth, the deposition rate of LaFeO$_3$ needs to be lowered down to 0.225 Å/min (14× smaller than 3.2 Å/min). However, using RF power below 40 W leads to unstable plasma and deposition rate for LaFeO$_3$.

To obtain high quality La$_x$Bi$_{1-x}$FeO$_3$ epitaxial films with a broad range of reliable La concentration, we employ a "digital doping" approach using off-axis co-sputtering. The BiFeO$_3$ plasma is continuously on throughout the whole film growth using a 50 W RF power with $r_{\text{BFO}} = 4.5$ Å/min. For each unit cell thickness growth of BiFeO$_3$ (4 Å, with a period of 53.1 seconds), the LaFeO$_3$ plasma is periodically turned on (45 W, $r_{\text{LFO}} = 3.2$ Å/min) for $t_{\text{on}}$ seconds, and then turned off for $t_{\text{off}} = 53.1 - t_{\text{on}}$ seconds by computer control of the RF power supply for LaFeO$_3$. For each La concentration $x$, $t_{\text{on}}$ can be calculated using equation $x = \frac{(t_{\text{on}}/60)*r_{\text{LFO}}}{(4\text{ Å})+(t_{\text{on}}/60)*r_{\text{LFO}}}$. As an example, to grow a La$_{0.1}$Bi$_{0.9}$FeO$_3$ film, we obtain $t_{\text{on}} = 8.3$ seconds and $t_{\text{off}} = 44.8$ seconds for LaFeO$_3$ growth during each 53.1-second cycle of BiFeO$_3$ deposition. It should be noted that this digital doping" mode of off-axis co-sputtering allows for growth of La$_x$Bi$_{1-x}$FeO$_3$ or other alloy-phase compounds with any concentration of $x$ from 0% to 100%. We have also tested reducing the digital doping cycle to 2 Å BiFeO$_3$ with a period of 26.6 seconds ($t_{\text{on}}$ and $t_{\text{off}}$ also reduced by half correspondingly), which resulted in essentially the same crystalline quality and ferroelectric behavior in the La$_x$Bi$_{1-x}$FeO$_3$ films. The results presented in this work are based on a period of 4 Å BiFeO$_3$ and 53.1 seconds.

XRD $2\theta/\omega$ and rocking curve scans were measured using a Malvern Panalytical Empyrean diffractometer with a 4-bounce Ge(110) monochromator to determine the crystalline quality of the La$_x$Bi$_{1-x}$FeO$_3$ (50 nm) films with $x$ ranging from 0.05 to 0.6 on SrTiO$_3$ and DyScO$_3$, as shown in Figs. 1a and 1b, respectively. A broad peak near $2\theta = 46°$ can be seen in some of the scans, which is attributed to the 10-nm SrRuO$_3$ (002) peak. For La$_x$Bi$_{1-x}$FeO$_3$ films on SrTiO$_3$, the La$_x$Bi$_{1-x}$FeO$_3$ (002) peak gradually shifts from 44.712° to 44.973° as $x$ varies from 0.05 to 0.35, corresponding to an out-of-plane (OOP) lattice constant of $c = 4.050$ Å to 4.028 Å. The large OOP lattice constant is due to the sizable compressive strain in the La$_x$Bi$_{1-x}$FeO$_3$ films considering the pseudo-cubic lattice constants of BiFeO$_3$ ($a_{pc}$ = 3.96 Å) and LaFeO$_3$ ($a_{pc}$ = 3.93 Å) on cubic SrTiO$_3$ ($a$ = 3.905 Å).  Interestingly, as $x$ increases slightly from 0.35 to 0.37, the La$_x$Bi$_{1-x}$FeO$_3$ (002) peak abruptly jumps from 44.973° to 45.424°, or $c$ = 4.028 Å to 3.990 Å, respectively.  For the La$_{0.6}$Bi$_{0.4}$FeO$_3$ film, the (002) shifts to 45.860°, corresponding to $c$ = 3.954 Å.

The La$_x$Bi$_{1-x}$FeO$_3$ films on DyScO$_3$ exhibit a similar behavior as shown in Fig. 1b.  The La$_x$Bi$_{1-x}$FeO$_3$ (002) peak shifts from 45.569° to 45.847° ($c$ = 3.978 Å to 3.955 Å) as $x$ increases from 0.05 to 0.35, after which, the peak suddenly jumps to 46.202° ($c$ = 3.926 Å) for $x$ = 0.37.  For all the La$_x$Bi$_{1-x}$FeO$_3$ films on DyScO$_3$, the La$_x$Bi$_{1-x}$FeO$_3$ (002) peaks are located substantially to the right of their counterparts on SrTiO$_3$, because the DyScO$_3$(110) surface has an almost square lattice with $a$ = 3.94 Å, resulting a much smaller compressive strain for BiFeO$_3$ and a tiny tensile strain for LaFeO$_3$.  As the La concentration increases, the Laue oscillations become clearer, indicating higher uniformity.  The insets in Fig. 1 show the XRD rocking curves for the $x$ = 0.3 films with narrow full-width-at-half-maximum (FWHM) of 0.012° to 0.018°, corroborating the high crystalline ordering.  Most importantly, the sudden change in the La$_x$Bi$_{1-x}$FeO$_3$ (002) peak position as $x$ increases slightly from 0.35 to 0.37 for both series on SrTiO$_3$ and DyScO$_3$ implies a phase transition from a BiFeO$_3$-dictated structure to a LaFeO$_3$-dictated structure, which coincides with the transition from ferroelectric to non-ferroelectric La$_x$Bi$_{1-x}$FeO$_3$ films as discussed later.

Scanning transmission electron microscopy (TEM) imaging was performed on two 50-nm La$_{0.1}$Bi$_{0.9}$FeO$_3$ films grown on SrRuO$_3$/SrTiO$_3$(001) and SrRuO$_3$/DyScO$_3$(110), as shown in Fig. 2.  Clear perovskite ordering in the La$_{0.1}$Bi$_{0.9}$FeO$_3$ and SrRuO$_3$ films as well as their epitaxial registry and sharp interfaces with the and SrTiO$_3$ and DyScO$_3$ substrates demonstrate the high crystal quality of La$_{0.1}$Bi$_{0.9}$FeO$_3$/SrRuO$_3$ bilayers.

To characterize ferroelectric properties of the two series of La$_x$Bi$_{1-x}$FeO$_3$ films at various La concentrations, we performed piezoresponse force microscopy (PFM) imaging and switching of the ferroelectric domains in those films using a Bruker Dimension Icon Atomic Force Microscope with a PointProbe Plus Electrostatic Force Microscopy (PPP-EFM) conductive tip from Nanosensors. Figure 3 shows the in-plane (IP) and OOP PFM images of as-grown ferroelectric domain patterns for BiFeO$_3$(100 nm) and La$_{0.15}$Bi$_{0.85}$FeO$_3$(50 nm) films grown on 10-nm SrRuO$_3$ on SrTiO$_3$(001) and DyScO$_3$(110). The OOP PFM images for the 4 samples show mostly a single domain contrast, indicating that the OOP FE polarization of the as-grown films primarily points in one preferred direction: downwards towards the SrRuO$_3$ bottom electrode.[28-30] Meanwhile, the IP PFM images of the 4 films exhibit clear contrast of two shades of domains. It is well known that the FE polarization of BiFeO$_3$ films is along one of the 8 <111> orientations.[2] For (001)-oriented BiFeO$_3$ films, the FE polarization is mostly along one of the 4 downward-pointing <111> directions, resulting in an almost uniform (downward) OOP polarization (see Figs. 3b, 3d, 3f, and 3h) and two-shade FE domain patterns in IP PFM images (see Figs. 3a, 3c, 3e, and 3g). It should be noted that the FE domain sizes of the La$_{0.15}$Bi$_{0.85}$FeO$_3$ films are significantly smaller than those of the BiFeO$_3$ films, accompanied by a transition from the characteristic striped-pattern in BiFeO$_3$ films to a more complex domain configuration in La$_{0.15}$Bi$_{0.85}$FeO$_3$. This evolution has been attributed to a weakening of the rhombohedral lattice distortion induced by La substitution as the polarization points away from the <111> axis and more along <112>.[22]

To investigate the effect of La-substitution on FE switching behavior and explore the upper limit of La concentration for La$_x$Bi$_{1-x}$FeO$_3$ films that remain ferroelectric at room temperature, we performed reversible FE domain switching measurements by applying a bias voltage between the PFM tip and the SrRuO$_3$ bottom electrode. We first used the internal voltage source in the Bruker

AFM system to switch the FE domains. Figure S1 in the Supplementary Materials shows an example of such measurement on a $La_{0.15}Bi_{0.85}FeO_3$(50 nm)/SrRuO$_3$(10 nm)/SrTiO$_3$(001) sample. A -10 V bias voltage is first applied over a 3 μm × 3 μm region to switch the OOP FE polarization from down to up, followed by +10 V bias voltage over a 1 μm × 1 μm area at the center of the larger square to switch the OOP polarization from up to down. It requires an apparent bias voltage of ±10 V to reversibly switch the FE domains using the internal voltage source. However, it is well known that the actual voltage drop between the PFM tip and the SrRuO$_3$ bottom electrode is considerably lower than 10 V because of voltage drops in other parts of the PFM circuit.

To gain better control of the FE switching process, we used an external Keithley 2400 sourcemeter to apply a voltage between the PFM tip and SrRuO$_3$ through a NanoScope signal access module. As a comparison, Fig. S2 in the Supplementary Materials shows the IP and OOP PFM images of FE domain switching of the same $La_{0.15}Bi_{0.85}FeO_3$(50 nm)/SrRuO$_3$(10 nm) films on SrTiO$_3$(001), with bias voltages of ±2 V, ±3 V, and ±4 V provided by the Keithley source meter. Clearly, a bias voltage of ±3 V is sufficient to induce nearly complete FE switching, which is dramatically lower than the ±10 V switching voltage needed for the internal source, although the actual switching voltage between the PFM tip and SrRuO$_3$ should be even smaller because there are likely still voltage drops in the NanoScope signal access module. To measure the accurate switching voltage across the FE layer, a capacitor structure with two conducting electrodes sandwiching the FE film is needed, which will be investigated in the future.

Using the external voltage source meter, we performed FE domain switching measurements on all the $La_xBi_{1-x}FeO_3$ films with $x = 0.05, 0.1, 0.15, 0.25, 0.3, 0.35, 0.37$, and 0.6 on SrTiO$_3$ and DyScO$_3$ substrates as shown in Fig. 4. Because OOP PFM images directly reflect the FE domain switching between up and down induced by an out-of-plane bias voltage, while the IP PFM images are at much less controlled states, Fig. 4 only shows the OOP images. The bias voltage for each sample was chosen to enable reversible and uniform FE domain switching with an OOP phase change of 180°, indicating a complete switching between the down and up polarization states. As can be seen in Fig. 4, for the two series of $La_xBi_{1-x}FeO_3$ films grown on both $SrTiO_3$ and $DyScO_3$, uniform FE domains are clearly observed for La concentration from 0.05 up to 0.35, which can be completely and reversibly switching by a voltage of ±3 V to ±5 V using an external voltage source. For $La_xBi_{1-x}FeO_3$ films at $x = 0.37$ and higher, no FE domains are observed using PFM, implying the disappearance of FE ordering at La substitution between 0.35 and 0.37. This sharp transition agrees well with the XRD scans in Fig. 1 where the $La_xBi_{1-x}FeO_3$ (002) peak jumps suddenly from $x = 0.35$ to 0.37 for both series on $SrTiO_3$ and $DyScO_3$. This suggests that the transition from FE phase to non-FE phase in the $La_xBi_{1-x}FeO_3$ films likely coincides with a phase transition from a $BiFeO_3$-dictated rhombohedral structure at $x \leq 0.35$ to a $LaFeO_3$-dictated orthorhombic structure at $x \geq 0.37$.

In conclusion, we investigated ferroelectric behavior of $La_xBi_{1-x}FeO_3$ epitaxial films grown on $SrTiO_3$ (001) and $DyScO_3$ (110) substrates using a "digital doping" method by off-axis co-sputtering with La concentration varying over a broad range of 0.05 to 0.6. XRD scans and STEM imaging confirm the high crystalline quality and sharp interfaces of the $La_xBi_{1-x}FeO_3$ films on a $SrRuO_3$ bottom electrode layer. PFM images demonstrate uniform and reversible FE domain switching with relatively small bias voltages of ±3 to ±5 V for the two series of $La_xBi_{1-x}FeO_3$ films with La concentration ranging from 0.05 up to 0.35 using an external voltage source meter. To date, this is the highest La substitution in $La_xBi_{1-x}FeO_3$ films that remain ferroelectric, offering opportunities for tuning ferroelectric and multiferroic properties in perovskite epitaxial films and heterostructures.

Acknowledgement: This work was primarily supported by the Intel Center of Ferro-Electrics for Energy Efficiency (COFEEE). XRD characterization was supported by the Department of Energy (DOE), Office of Science, Basic Energy Sciences, under Grant No. DE-SC0001304. The XRD and PFM measurements were performed at the OSU NanoSystems Laboratory which was partially supported by the Center for Emergent Materials: an NSF MRSEC under award number DMR-2011876. Electron microscopy was performed at the Center for Electron Microscopy and Analysis at OSU.

Data Availability Statement: The data that supports the findings of this study are available within the article and its supplementary material.

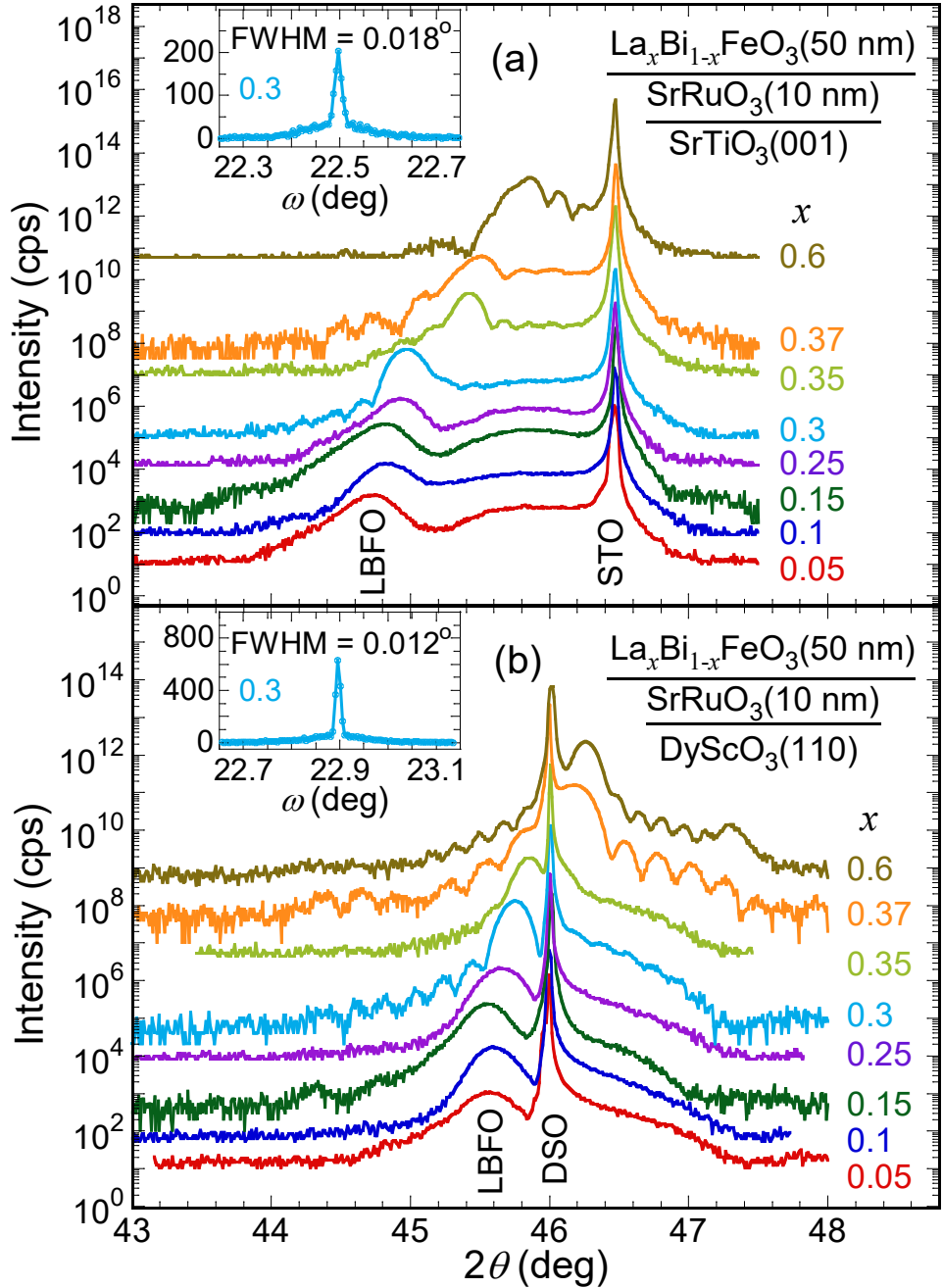

**Figure 1.** $2\theta/\omega$ XRD scans of La$_x$Bi$_{1-x}$FeO$_3$(50 nm) epitaxial films with a series of La doping concentrations from $x = 0\%$ to $60\%$ grown on an epitaxial SrRuO$_3$(10 nm) bottom electrode on (a) SrTiO$_3$(001) and (b) DyScO$_3$(110) substrates. As $x$ increases, the La$_x$Bi$_{1-x}$FeO$_3$(002) main peak gradually shifts to the right in general with increasingly clear Laue oscillations. A significant peak shift is observed between $x = 35\%$ and $37\%$, indicating a transition from a BiFeO$_3$ dominated structure to LaFeO$_3$ dominated one. The insets show the XRD rocking curves for the two films with 30% La, which exhibit narrow FWHM values of 0.012° and 0.018°.

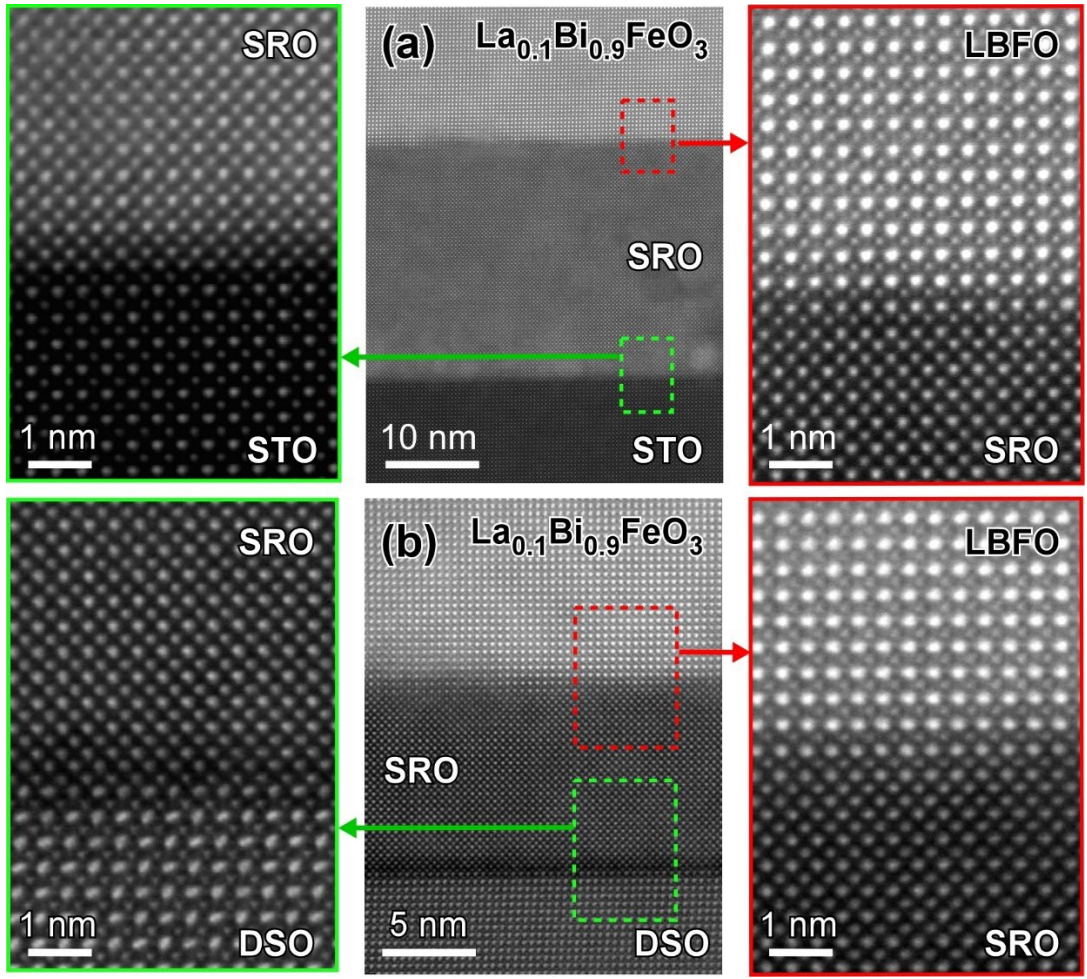

**Figure 2.** Cross-sectional STEM images of La$_{0.1}$Bi$_{0.9}$FeO$_3$ (50 nm) epitaxial films grown on (a) SrRuO$_3$(25 nm)/SrTiO$_3$(001) and (b) SrRuO$_3$(10 nm)/DyScO$_3$(110) viewed along the cubic <100> direction. The higher resolution STEM images of the SrRuO$_3$/SrTiO$_3$ and SrRuO$_3$/DyScO$_3$ interfaces (left) and the La$_{0.1}$Bi$_{0.9}$FeO$_3$/SrRuO$_3$ interfaces (right) reveal the sharp interfaces and clear epitaxial registry of the high-quality epitaxial layers.

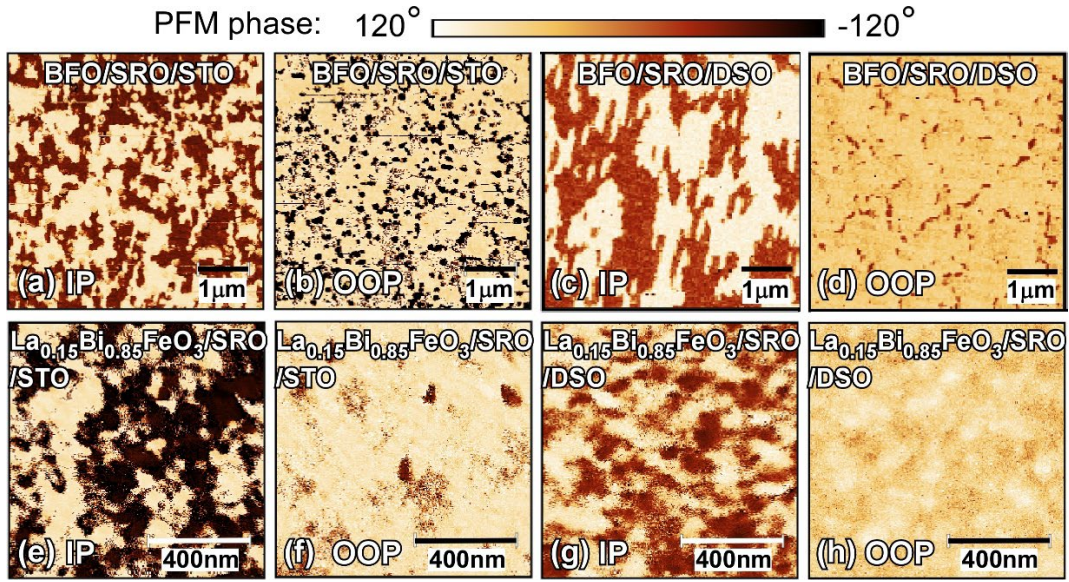

**Figure 3.** Initial ferroelectric domain patterns of BiFeO$_3$(100 nm)/SrRuO$_3$(10 nm) films grown on (a), (b) SrTiO$_3$(001) and (c), (d) DyScO$_3$(110), as well as La$_{0.15}$Bi$_{0.85}$FeO$_3$(50 nm)/SrRuO$_3$(10 nm) films grown on (e), (f) SrTiO$_3$(001) and (g), (h) DyScO$_3$(110), imaged by both in-plane (IP) and out-of-plane (OOP) PFM.

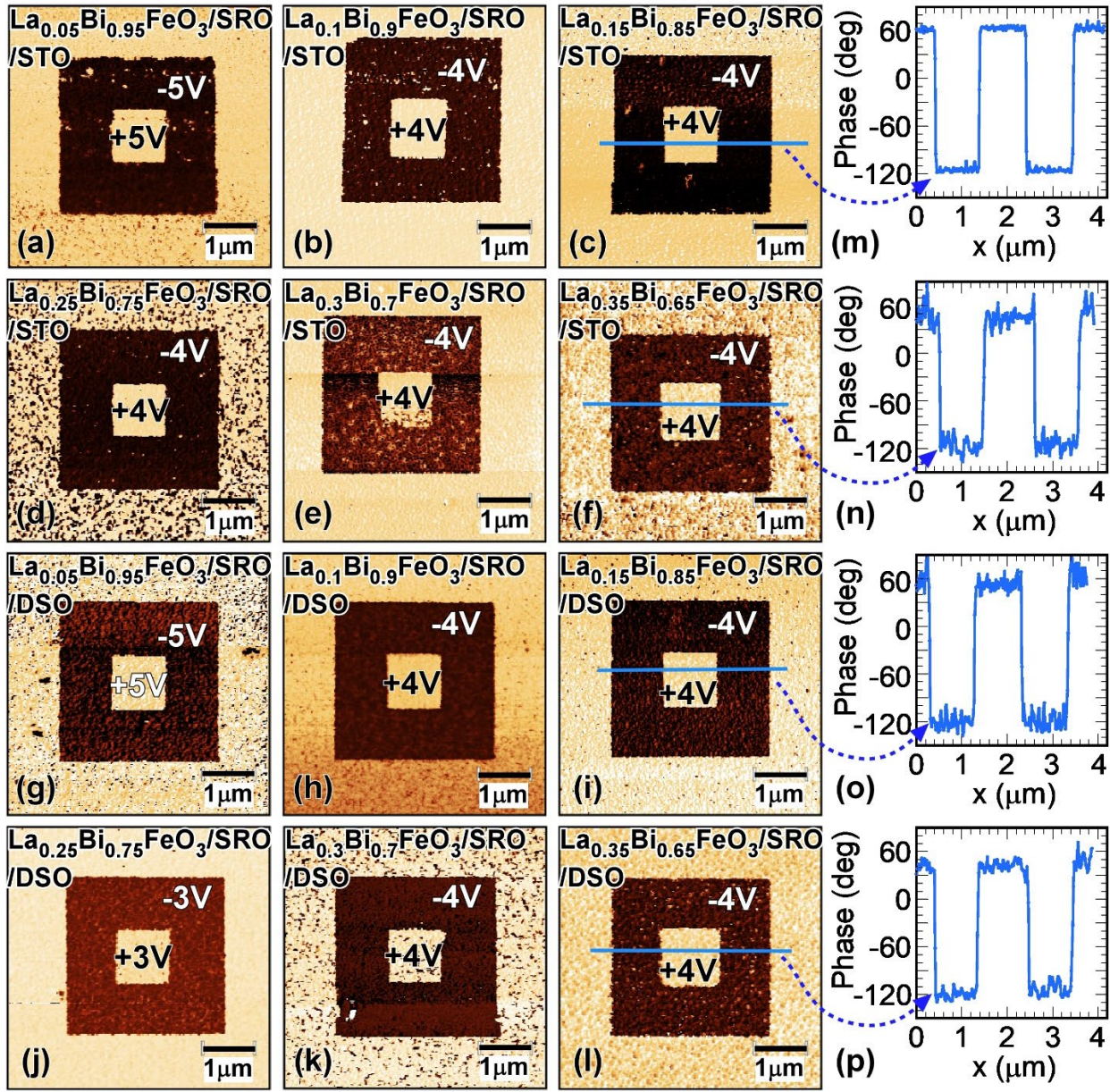

**Figure 4.** Switching of ferroelectric domains in La$_x$Bi$_{1-x}$FeO$_3$(50 nm)/SrRuO$_3$(10 nm) films with the La concentration ranging from 0.05 to 0.35 on (a)-(f) SrTiO$_3$(001) and (g)-(l) DyScO$_3$(110) substrates by out-of-plane PFM. A bias voltage of -3 to -5 V is first applied between the PFM tip and the SrRuO$_3$ bottom electrode over a 3 μm × 3 μm area to switch the FE polarization from up (bright) to down (dark) within the square, followed by PFM imaging. Then a bias voltage of +3 to +5 V is applied to a 1 μm × 1 μm area at the center of the larger square to switch back the FE polarization from down (dark) to up (bright). (m), (n), (o), (p) Linecuts of the out-of-plane PFM phase in the PFM images in (c), (f), (i), (l), respectively, showing 180° phase switching.